\documentclass[twocolumn,showpacs,prb,amsmath,amssymb]{revtex4}
\usepackage{graphicx}
\usepackage{dcolumn}
\usepackage{bm}
\usepackage{threeparttable}

\begin{document}

\title{Structure and stability of graphene nanoribbons in oxygen,
carbon dioxide, water, and ammonia}
\author{Ari P. Seitsonen}
\affiliation{IMPMC, Universit\'{e} Paris 6 et 7, CNRS, IPGP, 140 rue de
Lourmel, F-75015 Paris, France, and Physikalisch-Chemisches Institut der
Universit\"{a}t Z\"{u}rich, Winterthurerstrasse 190, CH-8057 Z\"{u}rich,
Switzerland}
\author{A. Marco Saitta, Tobias Wassmann, Michele Lazzeri,
and Francesco Mauri}
\affiliation{IMPMC, Universit\'{e} Paris 6 et 7, CNRS, IPGP, 140 rue de
Lourmel, F-75015 Paris, France}
\date{\today}

\begin{abstract}
We determine, by means of density functional theory, the stability
and the structure of graphene-nanoribbon (GNR) edges in presence
of molecules such as oxygen, water, ammonia, and carbon dioxide.
As in the case of hydrogen-terminated nanoribbons, we find that
the most stable armchair and zigzag configurations are
characterized by a non-metallic/non-magnetic nature, and are
compatible with Clar's sextet rules, well known in organic
chemistry. In particular, we predict that, at thermodynamic
equilibrium, neutral GNRs in oxygen-rich atmosphere should
preferentially be along the armchair direction, while
water-saturated GNRs should present zigzag edges. Our results
promise to be particularly useful to GNRs synthesis, since the
most recent and advanced experimental routes are most effective in
water and/or ammonia-containing solutions.
\end{abstract}

\maketitle

\section{Introduction}

Narrow graphene nanoribbons~\cite{journals/PhysRevB/Nakada1996a}
are carbon allotropes that promise to combine the high carrier
mobility of graphene~\cite{journals/Science/Novoselov2004} with a
semiconducting nature due to quantum confinement in the lateral
direction. The resulting systems are likely to be suitable for
direct application as channel in field-effect transistors and
other semiconducting devices. The design of GNRs, and more
generally of graphene samples with controlled edges, is crucial in
determining their electronic and transport properties. Among the
variety of reported experimental approaches, three routes,
intensely explored in the last couple of years, seem at the
frontier of current research: \emph{i)} lithographic
patterning~\cite{journals/PhysRevLett/Han2007a,journals/PhysicaE/Chen2007,Tapaszto08,Nemes10};
\emph{ii)} sonochemical
methods~\cite{journals/science/Li2008a,wang08}; \emph{iii)}
metallic nanoparticle chemical etching (unzipping) of
graphite/graphene
(nanotubes)~\cite{Datta08,Ci08,Kosynkin09,Jiao09,Campos09,Jiao10}.
Gaining control of the edge profiles and structures is of course
one of the most important issues to solve, somehow defining the
potential impact of these approaches. The observation of graphene
edges~\cite{Girit09,Jia09} and/or
GNR~\cite{journals/science/Li2008a,Campos09} profiles at
atomic-scale resolution indicates that the mentioned experimental
methods are capable of producing smooth, likely non-chiral
(\emph{i.e.} no mixing of zigzag and armchair direction), edges.
However, little is known on \emph{how} to achieve either zigzag or
armchair edges, and about the actual chemical termination of the
unsaturated carbon bonds.

In this regard, the huge number of theoretical works that recently
appeared in the literature on
GNRs~\cite{pisani07,journals/PhysRevB/Yamashiro2003a,
journals/Nature/Louie2006a,journal/ApplPhysLett/Kan2007a,
journals/PhysRevLett/Yazyev2008a,okada2008,koskinen-2008,Huang2008,Cervantes2008,Hod2007,Rama2010}
has almost exclusively focused on single-hydrogen saturation of
the edge carbon atoms. This was mainly due to the surprising
discovery of spin-polarized electronic states localized on the
edges of zigzag GNRs
~\cite{pisani07,journals/PhysRevB/Yamashiro2003a}, possibly
turning them into half-metals under high electric
fields~\cite{journals/Nature/Louie2006a,
journal/ApplPhysLett/Kan2007a}, and spurring great interest in
view of possible applications in future spintronics. However, we
have recently~\cite{Wassmann08,Wassmann10} pointed out that the
single-H-terminated zigzag GNRs, and thus all those fascinating
electronic and magnetic properties they exhibit, might only be
stable at very difficult experimental conditions.

The hydrogen case-study is presently extended to GNRs terminated
with other relevant atoms and
molecules~\cite{Lee09,Berashevich10,Wu09,Wu10,Zheng10,Yu10}. In
fact, as mentioned above, experiments are often performed in
aqueous or ammonia-containing solutions, or in oxidizing
conditions. More generally water and oxygen contamination (for
example in air) is extremely important in determining the
electronic properties of graphitic systems. For example, the
chemical reactions that induce the unzipping of carbon nanotubes
are thought to leave O-terminated zigzag GNRs as end
products~\cite{Kosynkin09}. In this work we thus report density
functional theory (DFT) calculations of the energetics and
structures of various GNR edges terminated with atmospheric
molecules, such as oxygen, water, and ammonia, as functions of the
respective chemical potentials. The work is organized as follows:
in section~\ref{sec:methods} we describe the theoretical method
and define the relevant thermodynamic quantities; in
section~\ref{sec:oxygen} we present our systematic study on
oxygen-terminated zigzag and armchair GNRs. In
section~\ref{sec:water} we report our results on selected GNRs in
presence of water and ammonia molecules; discussion on the
electronic and magnetic properties and comparisons with
experimental data are drawn in section~\ref{sec:discussion}, while
section~\ref{sec:conclusions} is devoted to conclusions and
perspectives.

\section{Method}\label{sec:methods}

DFT calculations were performed with the {\sc PWSCF} code of the
Quantum ESPRESSO suite~\cite{QE} in a
plane-wave/ultrasoft-pseudopotential approach, adopting the
Generalised Gradient Approximation (GGA) from Ref.~\cite{PBE} for
the exchange-correlation functional. We used a kinetic
energy/charge cut-off of 30/300 Ry. Edges were simulated within a
super-cell geometry using a vacuum layer of 9.5~{\AA} between two
edges and of 10.6~{\AA} between two graphene planes. $L$ was fixed
according to the lattice-constant of graphene (2.46~\AA) and
atomic positions were allowed to fully relax. In the case of
armchair edges, the width of the ribbons was about 22.4~{\AA} and
for zigzag ribbons about 28.9~{\AA}. Electronic integrations were
done using equidistant, consistent grids along the periodic
direction of the ribbon corresponding to 12 k-points in the
smallest unit cell. Denser grids have been occasionally used in
the presence of metallic/magnetic states. We limited our study to
the thermodynamic stability of neutral systems, and as such
kinetics and charge effects are not included. In fact, although
graphene edges and GNR can be obtained experimentally in charged
atmospheres, their technological interest resides in their
electronic properties in the neutral state, \emph{i.e.} at a Fermi
level as close as possible to the Dirac point of the graphene band
structure. We argue that GNRs and graphene edges obtained at high
doping level might be extremely reactive when brought back to
neutrality, and recombine to form more stable edge structures.

\section{Oxygen and carbon dioxide}\label{sec:oxygen}

\begin{figure*}[t]
\includegraphics[width=175mm]{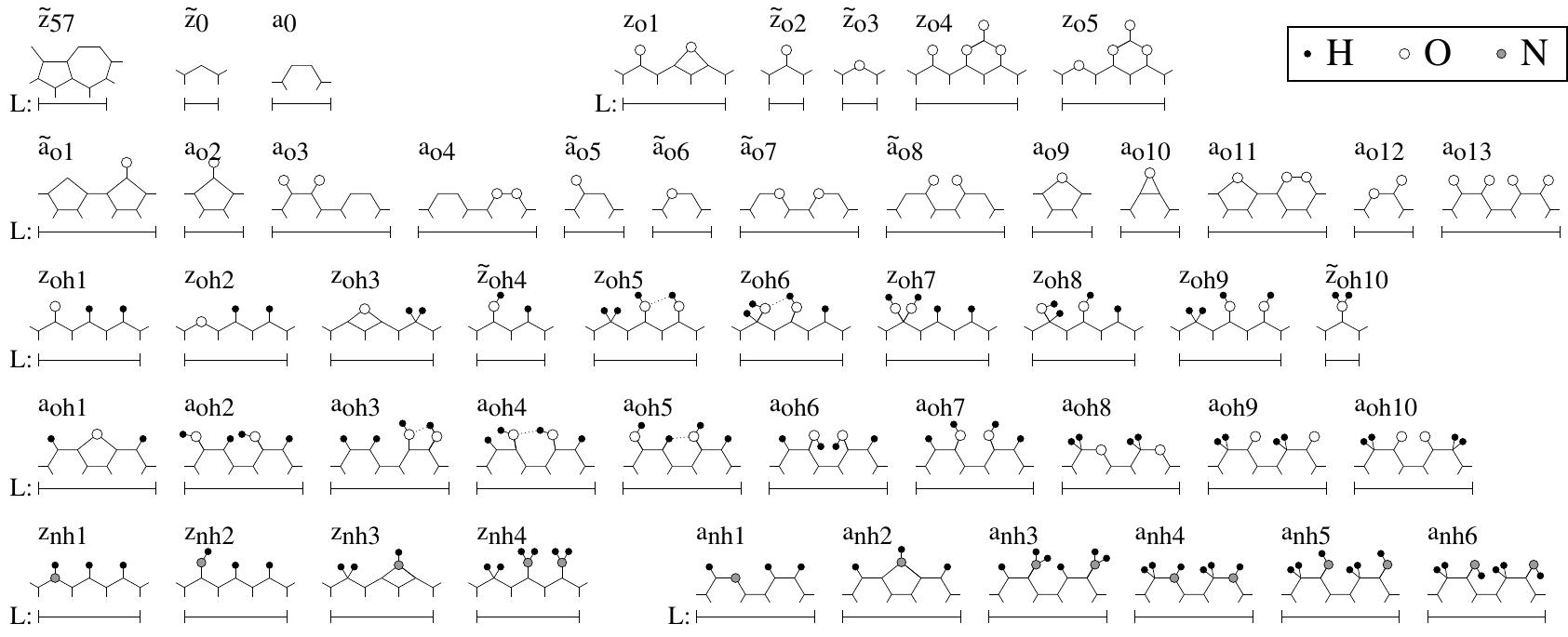}
\caption{ Scheme and labelling of the oxygen- (right top line and
second line from top), water- (third and fourth lines from top),
and ammonia-terminated (bottom line) edges of a graphene
nanoribbon considered in this work. Hydrogen, oxygen and nitrogen
atoms are the small circles, while carbon atoms are not explicitly
represented; dashed lines indicate hydrogen bonds. The structures
are periodic along the ribbon-edge with periodicity L, and are
ordered with respect to increasing O and N density and, at equal
density, to increasing formation energy. A tilde over an edge
label indicates that that structure is not an aromatic one.}
\label{fig:edgesall}
\end{figure*}

As in Ref.~\cite{Wassmann08}, we calculate the zero-temperature
edge formation-energy per unit length as follows:
\begin{equation}
{\cal E}_\mathrm{O_2} = \frac{1}{2L}\left( E^{ribb} - N_\mathrm{C}
E^{blk} -\frac{N_\mathrm{O}}{2}E_\mathrm{O_2}\right), \label{eq1}
\end{equation}
where $E^{ribb}$, $E^{blk}$ and $E_\mathrm{O_2}$ are the total
energy of the ribbon super-cell, of one atom in ``bulk'' graphene
and of the isolated O$_2$ molecule. $N_\mathrm{C}$
($N_\mathrm{O}$) are the number of carbon (oxygen) atoms in the
super-cell, while $L$ the length of the unit cell (see
Fig.~\ref{fig:edgesall}). ${\cal E}_\mathrm{O_2}$ can be used to
determine the stability of different structures as a function of
the experimental conditions~\cite{quian88}. In presence of
molecular O$_2$ gas, at a given chemical potential
$\mu_\mathrm{O_2}$, the relative stability is obtained by
comparing $G_\mathrm{O_2}={\cal E}_\mathrm{O_2}
-\rho_\mathrm{O}\mu_\mathrm{O_2}/2$, where
$\rho_\mathrm{O}=N_\mathrm{O}/(2L)$. At the absolute temperature
$T$ and for a partial O$_2$ pressure $P$~\cite{bailey07},
\begin{equation}
\mu_\mathrm{O_2} = H^\circ(T) - H^\circ(0) - TS^\circ(T) + k_BT
\ln\left(\frac{P}{P^\circ}\right), \label{eq2}
\end{equation}
where $H^\circ$ ($S^\circ$) is the enthalpy (entropy) at the
pressure $P^\circ = 1$~bar obtained from Ref.~\cite{H_data}. In
presence of monoatomic-oxygen gas one should use
$G_\mathrm{O}={\cal E}_\mathrm{O} -\rho_\mathrm{O}\mu_\mathrm{O}$,
where ${\cal E}_\mathrm{O}$ = ${\cal E}_\mathrm{O_2} -
\rho_\mathrm{O}\times 2.86$~eV, i.e. the binding energy \emph{per
atom} of the O$_2$ molecule in DFT-GGA. The most stable structures
are then obtained, as in Ref.~\cite{Wassmann08}, by comparing
their respective Gibbs free energy $G_\mathrm{O_2}$ as function of
the (molecular) oxygen chemical potential. $G$ is linear in $\mu$,
the slope being determined by $\rho_\mathrm{O}$. For a given value
of $\mu$ the most stable structure is the one with the lowest
value of $G$, thus, by increasing $\mu$ ({\it i.e.} going to an
environment richer in oxygen) the favorable structures will be
those with higher oxygen-density $\rho_\mathrm{O}$.

We report in Fig.~\ref{fig:edgesall} a pictorial representation of
the different GNR edges we have considered in this study, along
with their characteristic periodicity $L$. In the left top line we
show the clean zigzag and armchair GNRs~\cite{koskinen-2008}. The
right top line (second line from top) represent the zigzag
(armchair) oxygen-covered nanoribbons included in this work.
Structures are chosen as to saturate each oxygen atom with two
single bonds or a double one, while other possible chemical
bonding configurations are much less stable and are not included
in the results. The edges are ordered with respect to increasing
oxygen density and, at equal density, with respect to the
energetics, from most favorable to least favorable. Non-aromatic
edges~\cite{Wassmann08,Wassmann10} are indicated with a tilde (see
section~\ref{sec:discussion}).

\begin{figure}[ht!]
\includegraphics[width=80mm]{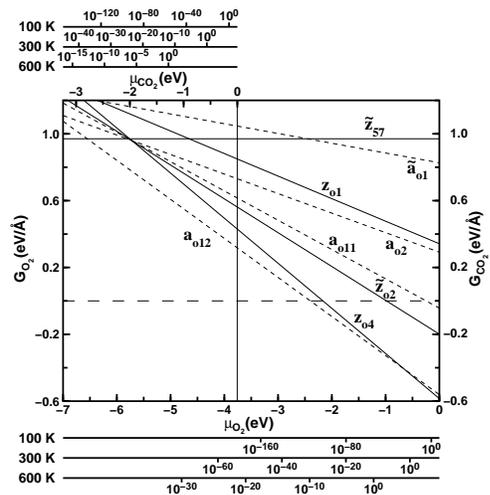}
\caption{Free energies versus chemical potential for the five most
stable oxygen-terminated edges, and of the clean one. The
alternative bottom (top) axes show the pressure, in bar, of
molecular O$_2$ (carbon dioxide) gas corresponding to the chemical
potentials at $\textrm{T}=100$, 300, and 600 K.}
\label{fig:edgeenergies}
\end{figure}

The density and zero-temperature edge-formation energy are
reported in Table~\ref{tab:1}. The free-energy G stability diagram
is reported in Fig.~\ref{fig:edgeenergies}. We note that three
zigzag GNRs, named $\tilde{\rm z}_{o2}$, z$_{o4}$, and z$_{o5}$,
and three armchair GNRs, named a$_{o11}$, a$_{o12}$, and
a$_{o13}$, have negative free energies, \emph{i.e.} graphene
should spontaneously break and form such ribbons if the
experimental conditions allow to overcome the reaction barriers,
as for example when nanoparticles, while sliding on graphene
layers, are able to cut edges along their
path~\cite{Datta08,Ci08}. We note that those experiments are
performed in hydrogen atmosphere; in the case of oxygen atmosphere
the reaction is more exothermic and could therefore be less
controllable.

The thermodynamic stability diagram (Fig. 2) indicates that the
armchair a$_{o12}$ and the zigzag z$_{o4}$ are the most stable
configurations at low and high concentrations of atomic or
molecular oxygen, respectively. However, the former one is more
stable for nearly all negative values of the O$_2$ chemical
potential, which suggests that the armchair edge should be the
only thermodynamically stable one in ordinary experimental
conditions going from ultra-vacuum to atmospheric concentration of
molecular oxygen.

On the other hand, it is even more important to consider the
presence of CO$_2$ molecules, either from the atmosphere or from
the oxidation of edge carbon atoms. We can define, analogously to
Eq.~\ref{eq1}, an edge formation energy per unit length with
respect to carbon dioxide, which reads:
\begin{equation}
{\cal E}_\mathrm{CO_2} = \frac{1}{2L}\left( E^{ribb} - \left(
N_\mathrm{C} - \frac{N_\mathrm{O}}{2} \right) E^{blk}
-\frac{N_\mathrm{O}}{2}E_\mathrm{CO_2}\right). \label{eq3}
\end{equation}

By simple calculations one can easily show that
\begin{equation}
{\cal E}_\mathrm{CO_2} = {\cal E}_\mathrm{O_2} +
\rho_\mathrm{O}\Delta E_\mathrm{CO_2}, \label{eq4}
\end{equation}
where $\rho_\mathrm{O}$ is, as above, the oxygen density per unit
length, and $\Delta E_\mathrm{CO_2}=\left(
E^{blk}+E_\mathrm{O_2}-E_\mathrm{CO_2}\right)$ is the formation
energy of molecular CO$_2$ from graphene and oxygen molecules, and
equals -3.76~eV in DFT-GGA. In other words, the chemical reaction
$\mathrm{C}^\mathrm{blk}+\mathrm{O}_2\rightarrow \mathrm{CO}_2$ is
strongly exothermic, as well known from the fact that ordinary
coal burns in oxygen once ignited. As a consequence, the
molecular-oxygen thermodynamic stability scale equally holds for
carbon dioxide, with a horizontal shift of $\Delta
E_\mathrm{CO_2}$. Conversely, the molecular oxygen thermodynamic
stability scale must include a vertical line at
$\mu_\mathrm{O_2}=\Delta E_\mathrm{CO_2}=-3.76 ~\mathrm{eV}$: at
higher oxygen densities no edge decoration can be
thermodynamically more stable than CO$_2$. This, again, implies
that only the armchair a$_{o13}$ edge is realistically observable
at thermodynamic equilibrium and oxygen concentration lower than
the one which would induce spontaneous ignition of graphene with
oxygen to produce CO$_2$.

\section{Water and ammonia}\label{sec:water}

\begin{figure}[ht!]
\includegraphics[width=80mm]{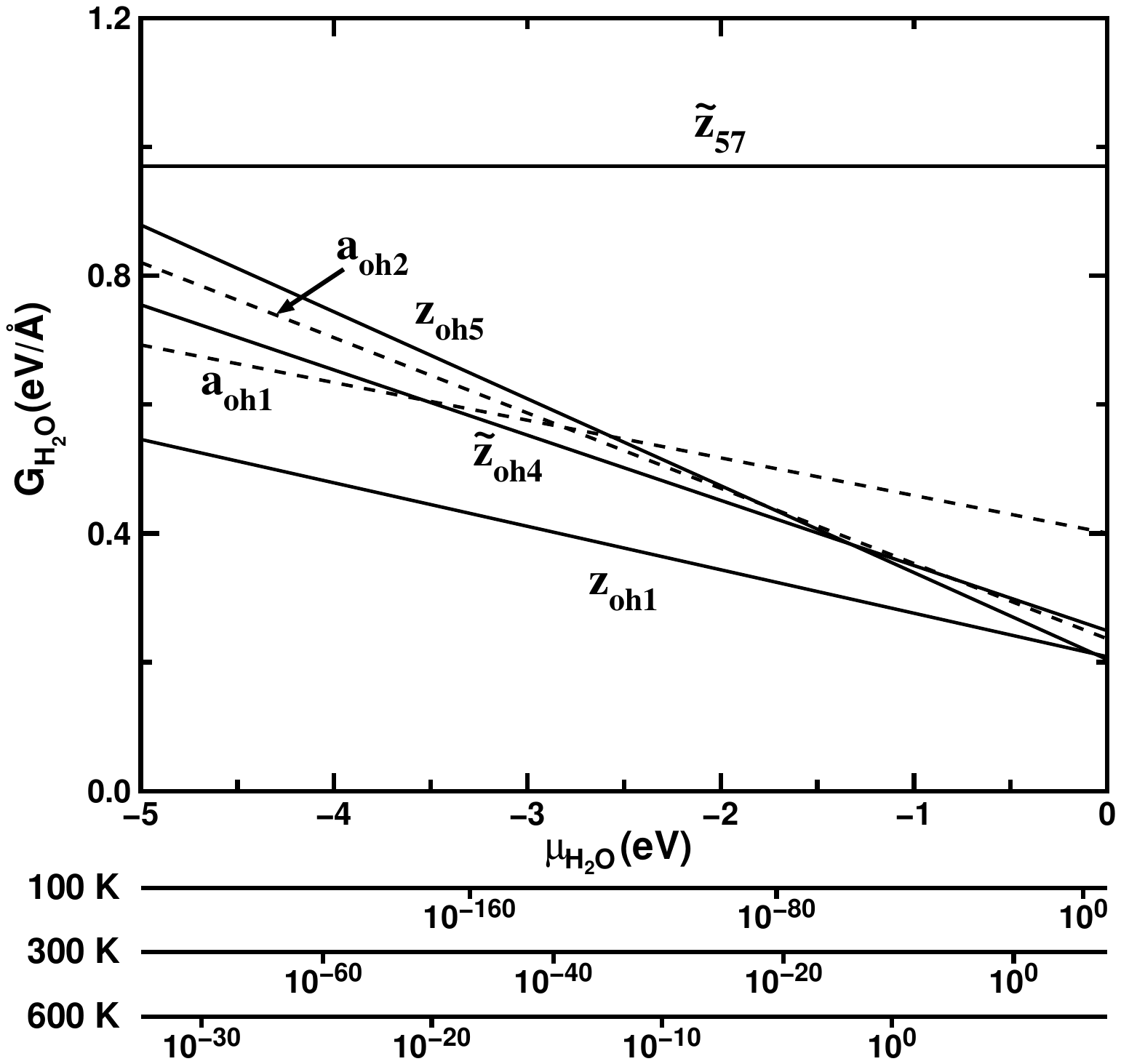}
\caption{Free energies versus chemical potential for the eight
most stable edges after exposure to water, including the clean
one. The bottom axes show the pressure, in bar, of molecular
H$_2$O vapor corresponding to the chemical potentials at
$\textrm{T}=100$, 300, and 600 K.} \label{fig:edgeenergiesH2O}
\end{figure}

\begin{figure}[ht!]
\includegraphics[width=80mm]{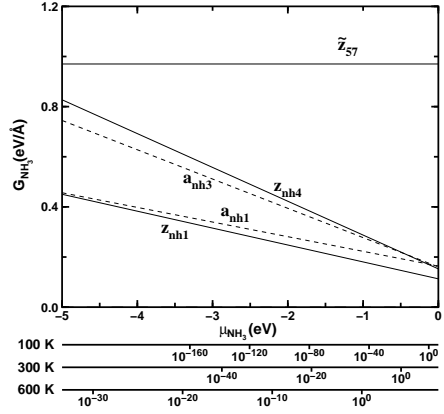}
\caption{Formation energies versus chemical potential for the four
most stable edges after exposure to ammonia, and of the clean one.
The bottom axes show the pressure, in bar, of molecular NH$_3$
vapor corresponding to the chemical potentials at
$\textrm{T}=100$, 300, and 600 K.} \label{fig:edgeenergiesNH3}
\end{figure}

Besides oxygen, we considered GNR edges saturated with water or
ammonia molecules. In these cases we restricted our study to O/H/N
contents as fixed by the respective molecular stoichiometries,
which corresponds to the description of experiments carried out in
pure water or ammonia atmospheres. In principle, the full study of
other stoichiometries would allow a more complete description of
the thermodynamic properties as function of two independent
chemical potentials, i.e. O and H (N and H) in the case of water
(ammonia). However, the number of different possible edges is
formidable, which would make these calculations extremely
demanding, virtually impossible to carry out ``by hand'', and
likely needing random-search optimizations~\cite{Pickard2006}.
Moreover, we only considered aromatic edge
configurations~\cite{Wassmann08}, with a few exceptions in the
case of water. The zero-temperature edge formation-energies per
length reads as follows:
\begin{equation}
{\cal E}_\mathrm{H_2O} = \frac{1}{2L}\left( E^{ribb} -
N_\mathrm{C} E^{blk} -N_\mathrm{H_2O}E_\mathrm{H_2O}\right)
\label{eq5}
\end{equation}
and
\begin{equation}
{\cal E}_\mathrm{NH_3} = \frac{1}{2L}\left( E^{ribb} -
N_\mathrm{C} E^{blk} -N_\mathrm{NH_3}E_\mathrm{NH_3}\right)
\label{eq6}
\end{equation}
in the case of water and ammonia, respectively. The restriction on
aromatic edges implies that zigzag edges respect the 2-1-1 bond
order rule for the dangling carbon bond that, as shown in
Refs.~\cite{Wassmann08,Wassmann10}, is the only aromatic
configuration for zigzag GNRs. In the case of water, single
dangling bonds are thus saturated by H atoms or OH groups, while
double bonds are saturated either by oxygen atoms, or by
H/OH-containing pairs (see Fig.~\ref{fig:edgesall}, bottom row).
In the case of ammonia, we considered all water-decorated
structures, and replaced each O atom (OH group), by a NH (NH$_2$)
group. The formation energies are summarized in table~\ref{tab:1},
the corresponding thermodynamic stability plots are shown in
Fig.~\ref{fig:edgeenergiesH2O} and Fig.~\ref{fig:edgeenergiesNH3}.

In the case of water, two main results are evident: the free
energy is always positive, meaning that graphene would not
spontaneously break to form edges in a water atmosphere. On the
other hand, any freshly-cut edge would be substantially stabilized
in the presence of water by the saturation of its dangling bonds
with O/H/OH groups. The other result is that z$_{oh1}$ is the most
stable configuration at any chemical potential. This means that,
in principle, the presence of oxygen favors the armchair edges,
while the presence of water favors the zigzag ones. In the case of
ammonia, free energies are always positive, and the z$_{nh1}$
zigzag GNR is the most stable one, in correspondence to z$_{oh1}$
in presence of water.

\section{Discussion}\label{sec:discussion}

We have shown in our previous works on hydrogen-decorated
edges~\cite{Wassmann08} and GNRs~\cite{Wassmann10} that a number
of different edge structures with different electronic/magnetic
properties compete for thermodynamic stability at different
hydrogen concentrations, and that they should be considered in
order to provide results of practical use to experimentalists. In
particular, we showed that an important hint on the structural and
chemical stability of GNR edges comes from the well-known organic
chemistry structural properties of polycyclic aromatic
hydrocarbons (PAHs), large molecules consisting of fused carbon
rings. According to the so-called Clar's
rule~\cite{books/Clar1964a,books/Clar1972a}, the most stable
structures for PAHs are the ones maximizing the number of Clar
sextets.

\begin{figure}[t]
\includegraphics[width=120mm]{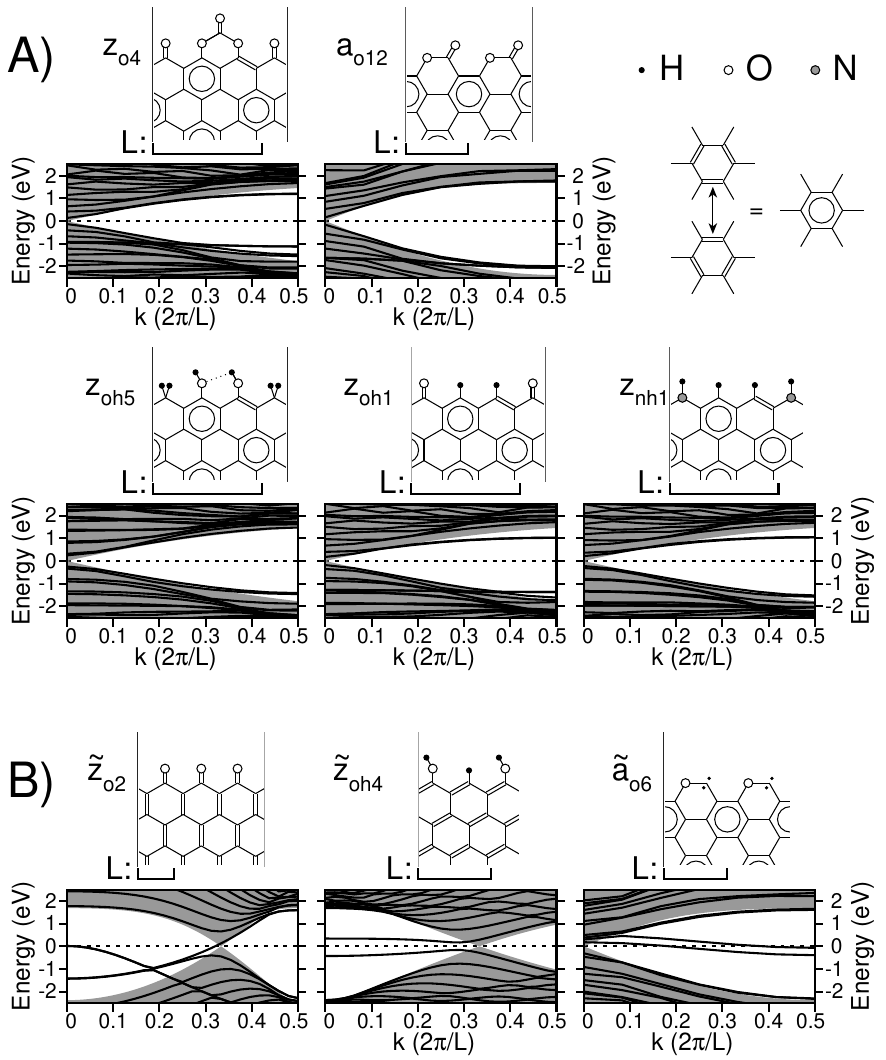}
\caption{ A) Scheme and electronic band-structure of the five most
stable oxygen-, water- and ammonia-terminated edges of a graphene
nanoribbon. Carbon-carbon bonds are represented with the standard
notation, while hydrogen, oxygen and nitrogen atoms are the small
circles. The structures are periodic along the ribbon-edge with
periodicity L. The gray area corresponds to the electronic-bands
allowed in ``bulk'' graphene. The dashed line is the Fermi level.
The top-right inset displays the standard representation of the
aromatic carbon ring. B) Three examples of "metallic/magnetic"
edges (see the definition in the text). $\tilde{z}_{o2}$,
non-magnetic within DFT-GGA calculations, is found to be magnetic
when hybrid DFT functionals are used~\cite{Rama2010};
$\tilde{z}_{oh4}$ and $\tilde{a}_{o6}$ both magnetic. The
$\tilde{a}_{o6}$ structure is non-aromatic according to our
definition, since the two small dots represent a diradical, which
is the only representation compatible with the carbon chemical
valence.} \label{fig:bands}
\end{figure}

In bulk graphene all $\pi$-electrons take part in the formation of
Clar sextets. In this arrangement, one out of 3 hexagons is a Clar
sextet. We define, as in Refs.~\cite{Wassmann08,Wassmann10},
aromatic edges as those allowing for the same 1/3 density of
benzenoid rings as bulk graphene, and in which every carbon atom
is fully saturated (it has 4 covalent bonds). Conversely, in
non-aromatic edges the 1/3 bulk-graphene-density of benzenoid
rings is incompatible with a full saturation of all carbon
dangling bonds. These edges are indicated with a ~$\tilde{ }$~ in
their label. This density can only be recovered at the price of
having carbon atoms with more or less than 4 covalent bonds. In
the presence of magnetism such over/under-bonding can be
interpreted as radicals, as in the edge labelled $\tilde{\rm
a}_{o6}$ (see Fig.~\ref{fig:bands}, panel B, and discussion in
Ref.~\cite{Wassmann10}).

As shown in Fig.~\ref{fig:edgesall} and in Table~\ref{tab:1}, we
find that, except for $\tilde{\rm z}_{o2}$ and analogously to the
case of hydrogen-saturated edges, the energetically most stable
ones are aromatic. In the case of water, the $\tilde{\rm z}_{oh4}$
edge is the only relatively stable non-aromatic configuration. In
Fig.~\ref{fig:bands} we report the electronic band structure,
close to the Fermi level, of the most stable and interesting
edges. The gray-shaded areas indicate the edge-projected bulk
bands. The full lines correspond to GNR electronic states. Bands
outside the gray-shaded area are electronic states exponentially
localized at the edges (edge states).

In line with the results of Ref.~\cite{Wassmann08}, in
Table~\ref{tab:1} and in Fig.~\ref{fig:bands} we show that
non-aromatic edges invariably display edge states at the Fermi
level that can split up and thus induce a magnetic character of
the edge. In absence of such splitting those states are only
partially occupied, giving rise to metallic behavior. On the
contrary, aromatic edges either do not possess edge states (as
shown in Fig.~\ref{fig:bands}, panel A) or such states are fully
empty or occupied (as ${\rm a}_{0}$~\cite{Wassmann08}).

\section{Conclusions}\label{sec:conclusions}

In conclusion, our ab initio DFT calculations indicate that
neutral graphene edges, in the presence of molecular species
typical of experimental conditions, are stabilized by decoration
with double-bonding O/NH groups or single bonding OH/H/NH$_2$
groups. In contrast to the case of hydrogen saturation, for each
molecular species we observe only very few possible edges in the
thermodynamical stability diagram, namely an armchair edge in the
case of oxygen, and one zigzag edge in the case of water and in
the case of ammonia. This observation might help experimentalists
in discriminating between zigzag and armchair edges and their
mixtures according to the chemical composition of the atmosphere
in which GNRs are obtained.

\begin{table}
\caption{Free energy ${\cal E}$ (defined as in
Eqs.~(\ref{eq1}),~(\ref{eq5}),~(\ref{eq6})), oxygen density
($\rho_\mathrm{O}=N_\mathrm{O}/(2L)$) and nitrogen density
($\rho_\mathrm{N}=N_\mathrm{N}/(2L)$) for all the studied edges.
The $\tilde{\rm z}_{o2}$ edge, non-magnetic within DFT-GGA
calculations, is found to be magnetic when hybrid DFT functionals
are used~\cite{Rama2010}.}
\begin{tabular*}{\columnwidth}{lcc|lcc|lcc}
 & $\rho_{\rm O}$     & ${\cal E}_{\rm O_2}$ &
 & $\rho_{\rm O}$     & ${\cal E}_{\rm H_2O}$ &
 & $\rho_{\rm N}$     & ${\cal E}_{\rm NH_3}$ \\
 & (\AA$^{-1}$)& (eV/\AA) &
 & (\AA$^{-1}$)& (eV/\AA) &
 & (\AA$^{-1}$)& (eV/\AA) \\
\hline
$\tilde{\rm z}_{57}^*$       & 0.000 &  0.9702 & $\tilde{\rm z}_{57}^*$        & 0.000 &  0.9702 & $\tilde{\rm z}_{57}^*$   & 0.000 &  0.9702 \\
$\tilde{\rm z}_0^\dagger$    & 0.000 &  1.1415 & $\tilde{\rm z}_0^\dagger$     & 0.000 &  1.1415 &$\tilde{\rm z}_0^\dagger$ & 0.000 &  1.1415 \\
   z$_{o1}$                  & 0.270 &  0.3420 &  z$_{oh1}$                    & 0.135 &  0.2086 &  z$_{nh1}$               & 0.135 &  0.1130 \\
$\tilde{\rm z}_{o2}^*$       & 0.405 & -0.1991 &  z$_{oh2}$                    & 0.135 &  0.2277 &  z$_{nh2}$               & 0.135 &  0.2291 \\
$\tilde{\rm z}_{o3}^*$       & 0.405 &  0.0570 &  z$_{oh3}$                    & 0.135 &  0.7640 &  z$_{nh3}$               & 0.135 &  0.6512 \\
   z$_{o4}$                  & 0.540 & -0.5847 & $\tilde{\rm z}_{oh4}^\dagger$ & 0.202 &  0.2364 &  z$_{nh4}$               & 0.270 &  0.1527 \\
   z$_{o5}$                  & 0.540 & -0.5723 &  z$_{oh5}$                    & 0.270 &  0.2042 &                          &       &         \\
                             &       &         &  z$_{oh6}$                    & 0.270 &  0.2235 &                          &       &         \\
                             &       &         &  z$_{oh7}$                    & 0.270 &  0.2518 &                          &       &         \\
                             &       &         &  z$_{oh8}$                    & 0.270 &  0.2614 &                          &       &         \\
                             &       &         &  z$_{oh9}$                    & 0.270 &  0.2619 &                          &       &         \\
                             &       &         & $\tilde{\rm z}_{oh10}^*$      & 0.405 &  1.4842 &                          &       &         \\
                             &       &         &                               &       &         &                          &       &         \\
    a$_{0}$                  & 0.000 &  0.9999 &    a$_{0}$                    & 0.000 &  0.9999 &    a$_{0}$               & 0.000 &  0.9999 \\
$\tilde{\rm a}_{o1}^\dagger$ & 0.117 &  0.8258 &  a$_{oh1}$                    & 0.117 &  0.4004 &  a$_{nh1}$               & 0.117 &  0.1643 \\
   a$_{o2}$                  & 0.234 &  0.2915 &  a$_{oh2}$                    & 0.234 &  0.2363 &  a$_{nh2}$               & 0.117 &  0.2808 \\
   a$_{o3}$                  & 0.234 &  0.3204 &  a$_{oh3}$                    & 0.234 &  0.2394 &  a$_{nh3}$               & 0.234 &  0.1605 \\
   a$_{o4}$                  & 0.234 &  0.3776 &  a$_{oh4}$                    & 0.234 &  0.2457 &  a$_{nh4}$               & 0.234 &  0.1880 \\
$\tilde{\rm a}_{o5}^\dagger$ & 0.234 &  0.4906 &  a$_{oh5}$                    & 0.234 &  0.2466 &  a$_{nh5}$               & 0.234 &  0.3186 \\
$\tilde{\rm a}_{o6}^\dagger$ & 0.234 &  0.4940 &  a$_{oh6}$                    & 0.234 &  0.2479 &  a$_{nh6}$               & 0.234 &  0.3310 \\
$\tilde{\rm a}_{o7}^\dagger$ & 0.234 &  0.5015 &  a$_{oh7}$                    & 0.234 &  0.2715 &                          &       &         \\
$\tilde{\rm a}_{o8}^\dagger$ & 0.234 &  0.5744 &  a$_{oh8}$                    & 0.234 &  0.3023 &                          &       &         \\
   a$_{o9}$                  & 0.234 &  0.5868 &  a$_{oh9}$                    & 0.234 &  0.3063 &                          &       &         \\
  a$_{o10}$                  & 0.234 &  0.8212 & a$_{oh10}$                    & 0.234 &  0.3244 &                          &       &         \\
  a$_{o11}$                  & 0.351 & -0.0432 &                               &       &         &                          &       &         \\
  a$_{o12}$                  & 0.467 & -0.5610 &                               &       &         &                          &       &         \\
  a$_{o13}$                  & 0.467 & -0.2239 &                               &       &         &                          &       &         \\
\end{tabular*}
\begin{tabular*}{\columnwidth}{rl}
$\tilde{~}$ & Non aromatic \\
$*$ & Metallic edges \\
$\dagger$ & Magnetic edges \\
\end{tabular*}
\label{tab:1}
\end{table}

Calculations were done at IDRIS (project n$^\circ$ 081202).

\end{document}